# Epitaxial growth of KTiOAsO$_4$ by pulsed laser deposition for nonlinear frequency conversion

Clavel Adrien[1], Salaün Mathieu[1], Bastard Lionel[2], Lepoittevin Christophe[1], Boulanger Benoît[1]

1: Univ. Grenoble Alpes, CNRS, Grenoble INP, Institut Néel, 38000 Grenoble, France

2: Univ. Grenoble Alpes, CNRS, Grenoble INP, CROMA, 38000 Grenoble, France

1. ABSTRACT

Twin-photons and triple-photons generations require pump laser intensities of several hundred MW.cm$^{-2}$. Micrometer-sized crystalline waveguides allow such interactions to be achieved at relatively low energies by taking advantage of the confinement of light. The materials selected to achieve these phenomena in the present work are from the titanyl phosphate family. The chosen architecture is a micrometric layer of KTiOAsO$_4$ (KTA) grown on a KTiOPO$_4$ (KTP) substrate forming a planar waveguide in a first step. These two isostructural materials have a low lattice parameter mismatch (around 2 %), making it possible to achieve high-quality epitaxial growth. The refractive index of KTA being higher than that of KTP ($\Delta n \approx 0.043$ at 1500 nm), this couple of materials could allow light to be guided in the deposited layer. To grow the KTA film, the chosen technique is pulsed laser deposition (PLD) that allows nonlinear complex oxides layer of optical quality to be grown. In this study, we show that the growth is well oriented for layers with a thickness up to 200 nm after deposition under 32 mTorr of oxygen and a laser fluence of 0.8 J.cm$^{-2}$, forming epitaxial layers. Then, after annealing at 650°C for 14h, the crystal state is improved as well as the surface. However, our calculations show that the layer's thickness has to be equal to 1.68 μm to achieve SHG at Telecom wavelengths, which is the next step of this research. In this study we demonstrate the feasibility of KTA epitaxy over KTP by Pulsed Laser Deposition (PLD) with good epitaxial quality, good orientation and theoretical phase-matching conditions for a planar KTA waveguide.

2. INTRODUCTION

The development of nonlinear optics, with the generation of entangled twin-photons or triple-photons, opens up new opportunities in quantum information [1]. However, these applications often require the ability to work at low laser energy and in the Continuous Wave regime. The efficiency of nonlinear processes depends on the intensity, *i.e.* the energy *per* time and surface units, so that it is therefore important to spatially confine the pump laser in order to maximize the nonlinear conversion efficiency. Micrometric or sub-micrometric waveguides are therefore good candidates. There are various techniques for manufacturing waveguides in bulk nonlinear crystals, such as proton exchange[2], ion implantation [3] or dicing [4]. A good alternative to these techniques is to grow the crystal as a thin film using a method such as pulsed laser deposition (PLD). This technic is well suited for growing complex oxides thin films in a short amount of time[5]. PLD consists of the ablation by a pulsed high energy laser of a material of the layer's desired composition. The ablation forms a plasma plume which condenses on the substrate forming a layer with the same composition than the target[6]. The PLD technic has a high tendency to form directly an epitaxial layer during the deposition, nevertheless, a post-annealing can improve the crystal quality of the film. Deposition of nonlinear crystals by PLD has already been reported[7–11].

The work carried out here follows an initial study based on the PLD growth of RbTiOPO$_4$ (RTP) on KTiOPO$_4$ (KTP)[12]. In this study, it was shown that annealing led to interdiffusion of K$^+$ and Rb$^+$ ions, making it impossible to use the RTP layer as a waveguide with this growing process. RTP material was therefore abandoned in favor of KTiOAsO$_4$ (KTA), a material with nonlinear optical properties[13], which is also isomorphic to KTP with the same alkali. Since arsenic is engaged in As-O bonds and phosphorous in P-O bonds interdiffusion is avoided. Furthermore, KTA has lattice parameters close enough to those of KTP, ensuring epitaxial deposition and a large enough refractive index difference, allowing strong optical confinement.

The present work begins with numerical simulations on phase-matched second-harmonic generation (SHG) in a planar optical waveguide made of a KTA thin film deposited onto a KTP substrate, which allows us to define a roadmap for the elaboration process described in the following. Then we define the relevant deposition parameters in order to obtain a dense layer of KTA, followed by the description of the different characterizations of the final layer.

KTA, as KTP, is a positive biaxial crystal so that the relation of order between the three refractive indices is: $n_x(\lambda) < n_y(\lambda) < n_z(\lambda)$ where $\lambda$ is the wavelength and the Cartesian indices refer to the dielectric frame $(x, y, z)$ that is collinear to the crystallographic frame $(\boldsymbol{a}, \boldsymbol{b}, \boldsymbol{c})$[14]. Except along the two optical axes of the index surface, any direction of propagation exhibits birefringence, which means that two possible refractive indices can be excited according to the polarization of light. This property gives the possibility to achieve phase-matching between the nonlinear polarization and the radiated field during the phenomena of sum-frequency generation (SFG: $\frac{1}{\lambda_1} + \frac{1}{\lambda_2} = \frac{1}{\lambda_3}$) or difference-frequency generation (DFG: $\frac{1}{\lambda_1} - \frac{1}{\lambda_2} = \frac{1}{\lambda_3}$), which leads to a maximization of the conversion efficiency. Phase-matching is then an important target that has to be taken into account in the framework of materials engineering when the goal is to fabricate efficient nonlinear media. We are interested here in second-harmonic generation (SHG), which is a particular case of SFG where $\lambda_1 = \lambda_2 \ (\equiv \lambda_\omega)$ and so $\lambda_3 = \frac{\lambda_\omega}{2} \ (\equiv \lambda_{2\omega})$. And because propagation is envisaged in a waveguide, the refractive index has to be replaced by the effective index, $n_{eff}$, in order to take-into-account the confinement geometry, in particular the thickness $e$ of the layer[15]. Then in a first step, it is necessary to calculate the magnitude of the three principal effective indices, $n_{eff}^x$, $n_{eff}^y$ and $n_{eff}^z$, as a function of wavelength, starting from the dispersion equations of the three principal refractive indices[14]. Note that $n_{eff}^x(\lambda) < n_{eff}^y(\lambda) < n_{eff}^z(\lambda)$. The planar waveguide dispersion equation in an anisotropic material can be simplified by aligning the direction of propagation of the guided mode with one axes of the dielectric frame of the KTA layer. An analytical mode solver for the 3-layer system (KTP, KTA, air) can then be used to compute the effective index of propagation $n_{eff}$ for the two possible directions of the electric field.

Using these equations of dispersion, we identified the y-axis as the direction of propagation allowing to reach type II phase-matching. The corresponding relation writes:

$$2n_{eff}^x(\lambda_{2\omega}, e) = n_{eff}^x(\lambda_\omega, e) + n_{eff}^z(\lambda_\omega, e) \qquad (1)$$

Equation (1) shows that the two orthogonally polarized modes, along the x- and z-axis, are involved at the fundamental wavelength $\lambda_\omega$. It requires to properly polarize the pump beam, *i.e.* at 45° of the x- and z-axis, as shown in Figure 1 (top), respectively corresponding to the transverse magnetic (TM) and transverse electric (TE) polarizations in term of guided optics. In Figure 1 (bottom) is plotted the phase-matching curve given the fundamental wavelength $\lambda_\omega$ as a function of the thickness $e$ of the KTA layer.

The curve of Figure 1 shows that phase-matching is not possible for a layer thickness smaller than $e = 1.62$ µm, corresponding to $\lambda_\omega = 1.70$ µm. Above this minimal thickness, phase-matching is possible for two fundamental wavelengths: for example, at $e = 1.68$ µm correspond $\lambda_\omega = 1.50$ µm and $\lambda_\omega = 1.88$ µm. For such a layer thickness, the waveguide is single-mode at the fundamental wavelength and bi-mode at the second harmonic frequency.

Then the target for the frequency doubling at Telecom wavelengths is a layer thickness around 1.7 µm.

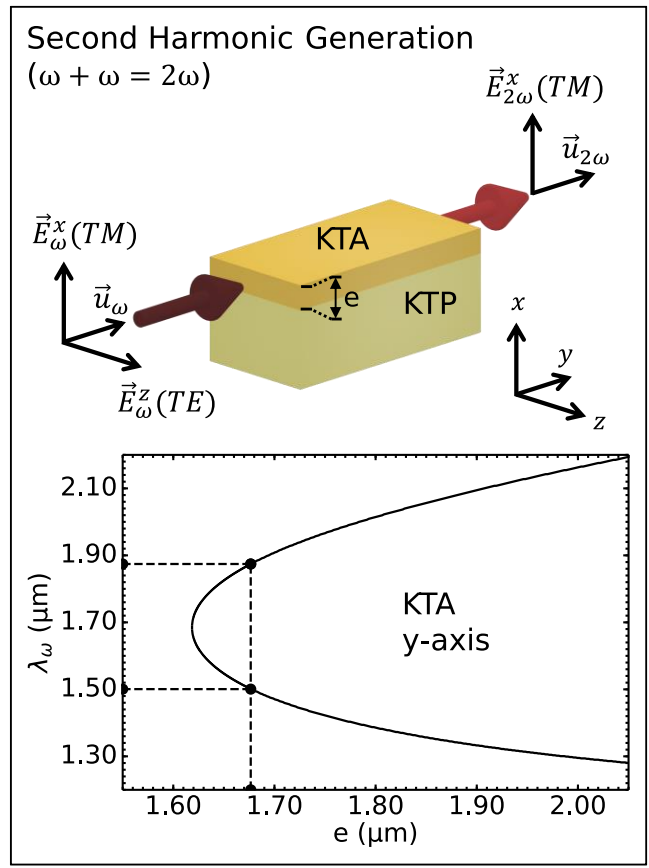

Figure 1 : Birefringence phase-matching curve calculated for a propagation along the y-axis of KTA and giving the fundamental wavelength $\lambda_\omega$ as a function of the thickness $e$ of the KTA layer; $(x, y, z)$ is the dielectric frame of KTA; $(\vec{E}_\omega^x(TM), \vec{E}_\omega^z(TE), \vec{E}_{2\omega}^x(TM))$ stand for the fundamental

and second-harmonic electric field polarizations; $(\vec{u}_\omega, \vec{u}_{2\omega})$ are the unit vectors of the fundamental and second-harmonic wave vectors.

In this work the deposition is done on X faced KTP ((h00) planes parallel to the surface) substrates grown by CRISTAL LASER SA then orientally cut and polished to a roughness of 4 Å at Institut Néel. Substrates are cleaned using deionized water, isopropanol and acetone baths, then fixed by a clamp on the substrate holder. To finish the cleaning process the substrate is placed in the deposition chamber and heated at 700°C under the ultimate vacuum of the chamber (around 3 µTorrs) for 1 hour. During deposition, the substrate is heated at 700°C and dynamic $O_2$ pressure inside the chamber is adjusted at 32 mTorr monitored by a Balzers Compact Full Range Gauge. The deposition is carried out by ablation of a KTA single crystal target using a KrF excimer laser (COHERENT COMPexPro 102 F) (246 nm, 20 ns pulses, 15Hz). The laser fluence on the target is set at 0.8 J.cm$^{-2}$ to reduce photochromic damage of the target, which is called "grey tracking"[16,17], by local heating due to the laser. To avoid ballistic particles deposition, a 1.05-cm-wide alumina mask is placed halfway between the target and the substrate. Finally, the samples are annealed during 14h in a tubular furnace under oxygen at 650°C to ensure good oxygen stoichiometry of the layer and good crystallization of the phase.

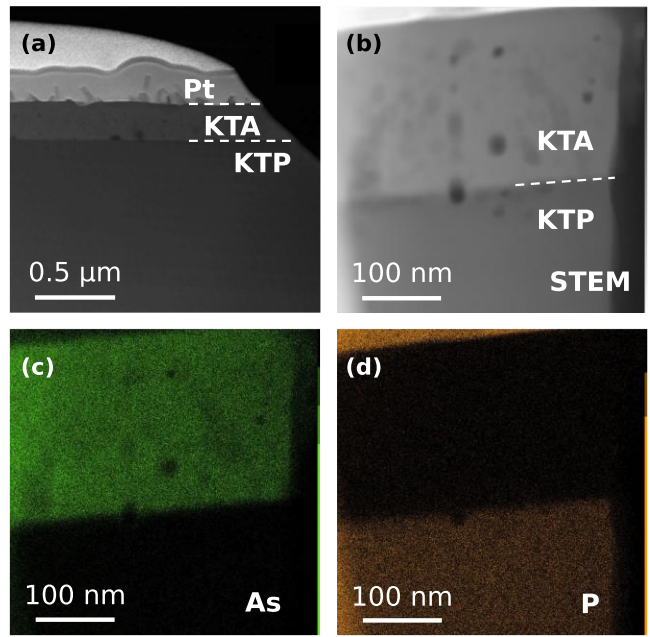

Figure 2 : (a) Low magnification STEM image of the film observed in cross section; (b) STEM image of the KTA layer over the KTP substrate, (platinum is seen in the upper left corner); (c) and (d) EDX elemental mapping of arsenic and phosphorus respectively. On (d), Platinum is observed due to convolution of the P and Pt peaks.

The cross-section TEM preparation was carried out by FIB, which explain the presence of a platinum layer above the KTA deposit observed on the low magnification STEM image shown in Figure 2a. Figure 2b allows to estimate the thickness of the layer around 215 nm, and also to confirm the good thickness homogeneity of the KTA layer. The dark spots observed in Figure 2b are due to unfortunate carbon contamination that occurred during STEM observation. The chemical composition of the sample was verified by performing EDX mapping over the whole area of this STEM image. The arsenic EDX signal (Figure 2c) is clearly observed in the KTA film area only, while phosphorous is present exclusively in the substrate region (Figure 2d). The comparison of the As and P EDX elemental mappings defines a straight interface between the substrate and the KTA film, with no interdiffusion, in contrast with the behavior seen during RTP film growth[12].

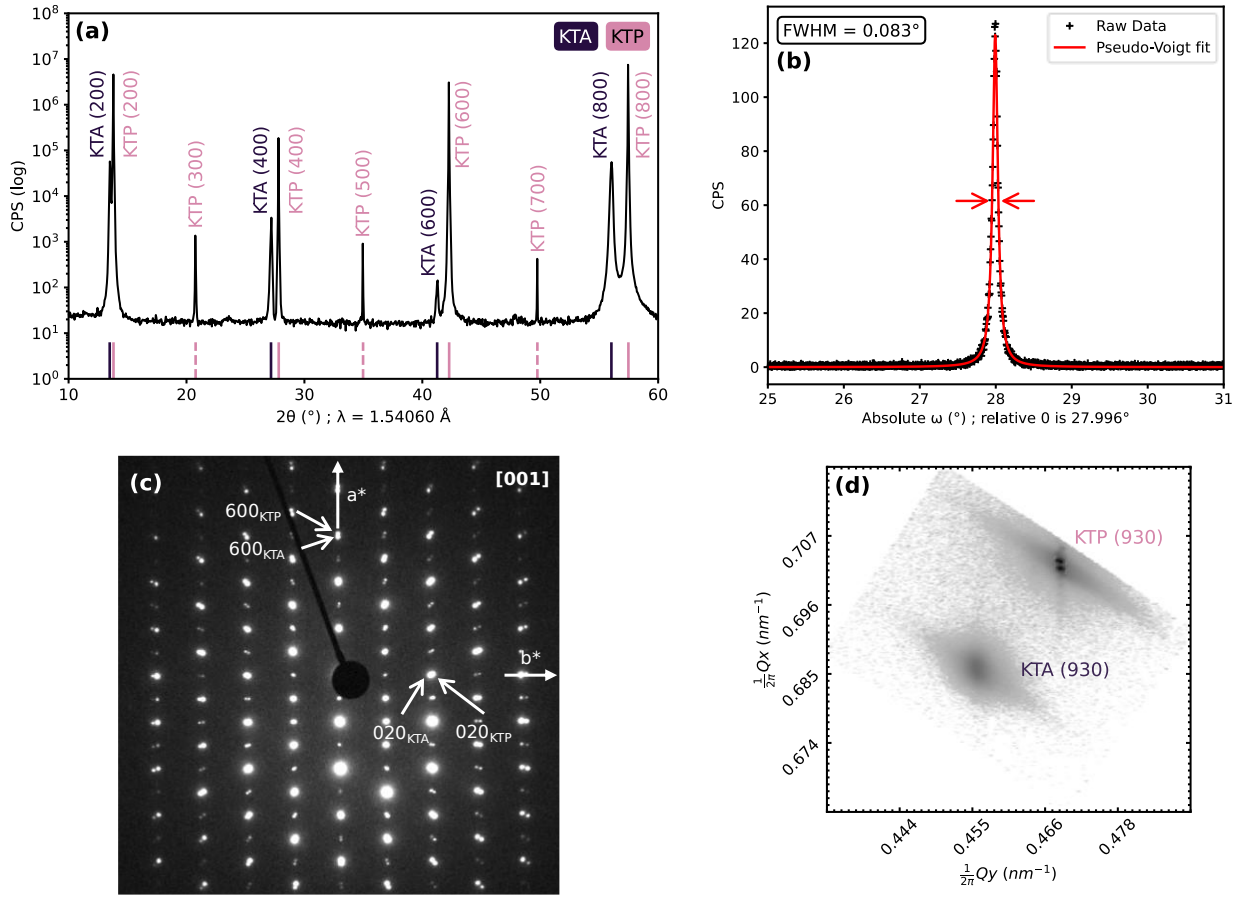

Figure 3 : Diffraction diagrams with (a) θ-2θ scan of the film where continuous lines correspond to PDF data (KTA: 04-009-2944; KTP: 01-078-1342) and the dashed lines to calculation; (b) ω-scan of the KTA (*800*) reflection; (c) [001] zone axis ED pattern of the interface between KTA film and KTP substrate; (d) Reciprocal Space Mapping (RSM) on the (*930*) reflections of KTA and KTP.

In addition to the stoichiometry and morphological studies, the diffractions analyses are valuable to asses not only the layer structure we are dealing with but also the crystal quality and strains of this layer. On the θ-2θ scan diffraction pattern (Figure 3a), the absence of peaks that are not KTA and KTP (*h00*) shows: *(i)* an oriented crystallization along the [h00] axis, *(ii)* that the layer is single crystalline, and *(iii)* that the composition of the target is well transferred on the substrate. The cell parameters of KTA and KTP calculated from this diagram are 13.11 ± 0.03 Å and 12.82 ± 0.03 Å, respectively: it corresponds to $\varepsilon_a \approx$ -0.178 ± 0.006 % in term of strain along the a-axis for the film which is a weak value. It should be noted that (*h00*) reflections where *h* are odd should be extinct. Their presence on the diffraction pattern is attributed to multiple diffraction due to the high crystal quality of the KTP substrate.

The KTA film's crystal quality has been evaluated by ω-scan on the (*800*) reflection (Figure 3b). The calculated full width at half maximum (FWHM) is 0.083 ± 0.001°. In comparison, KTP substrate (high quality flux grown single crystal) FWHM in the same measurement conditions is 0.005 ± 0.001°. It is worth noting that the intensity is different between Figure 3a and 3b: it comes from the fact that the monochromators used for the measurements were different, *i.e.* respectively 2×*(220)* Ge and 4×*(220)* Ge, the latter one decreasing the diffracted intensity by one order of magnitude.

The ED pattern oriented along the [001] direction (Figure 3c) was recorded at the interface between the film and the substrate to visualize their respective lattices. Indexation of the diffraction spots is consistent with the orthorhombic space group *Pna2₁*. As for Figure 3a, (*h00*) and (*0k0*) reflections, where *h* and *k* are odd, should be extinct, which is not the case due to multiple diffraction. RSM was done on (*802*) (SI 2) and (*930*) reflections (Figure 3d). From the spot's projections on Qx, Qy and Qz (see SI 2 for Qz) axis the calculated cell parameters are $a_{film} \approx$ 13.12

± 0.09 Å, $b_{film}$ ≈ 6.583 ± 0.005 Å, $c_{film}$ ≈ 10.75 ± 0.02 Å and $a_{substrate}$ ≈ 12.82 ± 0.09 Å, $b_{substrate}$ ≈ 6.404 ± 0.005 Å, $c_{substrate}$ ≈ 10.59 ± 0.02 Å. These values lead to strains along the layer axes of $\varepsilon_a$ ≈ -0.145 ± 0.006 %, $\varepsilon_b$ ≈ 0.041 ± 0.008 %, $\varepsilon_c$ ≈ 0.057 ± 0.007 %, which confirms the negative strain along the a-axis calculated from the θ-2θ scan, and positive strains along the b- and c-axes. We can conclude that the film is nearly relaxed in the plan without any epitaxial stress.

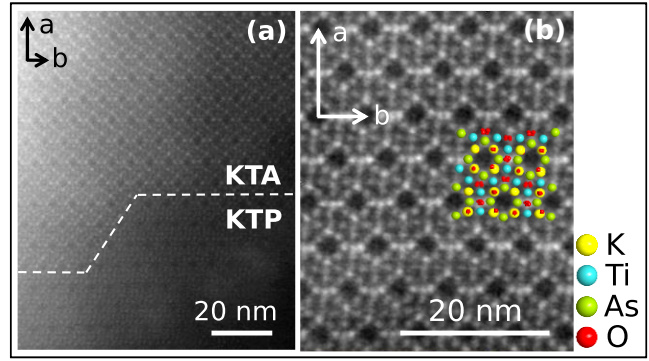

Figure 4 : [001] HAADF-STEM image of (a) KTA film deposited on KTP substrate and (b) KTA film with the corresponding orientation of the KTA crystal structure superimposed on it. The distortion is due to a slight shift during the scanning at high magnification.

Figure 4 exhibits HAADF-STEM images recorded along the [001] direction. In these images, the intensity of each projected ion column is directly proportional to the average atomic number denoted as $Z^n$ (1 < n < 2). As a result, the brightest dots correspond to the elements with the highest atomic numbers, with As (Z = 33) being the brightest, followed by Ti (Z = 22), K (Z = 19), and finally O (Z = 8). Figure 4a provides an overall view of the film deposited on the substrate, exhibiting perfect crystallinity without any visible crystal defects. It also evidences the flawless epitaxy of the film on the substrate, with a clean interface between the two. The interface appears stepped rather than perfectly straight, which may be attributed to the substrate preparation process. Figure 4b displays an enlarged HAADF-STEM image of the film, overlaid with the projected structural model of KTA. This comparison clearly shows the excellent agreement between the model and the contrast observed in the images. This confirms at small scale the chemical and structural analyses hereinbefore presented.

In conclusion, this work provides a theoretical and experimental corpus for the development of integrated optical devices allowing twin-photons or triple-photons generation. The simulation indicates that a 1.7 µm thick KTA layer would be able to achieve twin-photons generation in the Telecom range. We also demonstrated that the growth of an epitaxial 200 nm tick KTA layer can be obtained by pulsed layer deposition. The composition uniformity has been verified by EDX analysis and assed the absence of interdiffusion of As and P atoms. Diffraction analysis confirmed the Pna2$_1$ structure and showed the single crystalline state of the layer, as well as the small tilt and light strains. HAADF-STEM exhibited the high-quality epitaxy of the layer, while the absence of As diffusion and the atomic fit on the structural KTA model confirms the crystal orientation and structure.

ASSOCIATED CONTENT

Supporting Information: Characterization methods and additional experimental results (ED unit cell determination, KTA relaxed unit cell determination and RSM on (*802*) reflection).

AUTHOR INFORMATION

**Corresponding Author**

Salaün Mathieu − Univ. Grenoble Alpes, CNRS, Grenoble INP, Institut Néel, 38000 Grenoble, France; https://orcid.org/0000-0002-1513-9000 ; Email: mathieu.salaun@neel.cnrs.fr

Boulanger Benoît − Univ. Grenoble Alpes, CNRS, Grenoble INP, Institut Néel, 38000 Grenoble, France; https://orcid.org/0000-0003-2682-3913 ; Email: benoit.boulanger@neel.cnrs.fr

**Author**

Clavel Adrien − Univ. Grenoble Alpes, CNRS, Grenoble INP, Institut Néel, 38000 Grenoble, France; https://orcid.org/0009-0004-7703-1048.

Bastard Lionel − Univ. Grenoble Alpes, CNRS, Grenoble INP, CROMA, 38000 Grenoble, France; https://orcid.org/0000-0002-8121-9412.

Lepoittevin Christophe − Univ. Grenoble Alpes, CNRS, Grenoble INP, Institut Néel, 38000 Grenoble, France; https://orcid.org/0000-0002-3995-0746.

**Author Contributions**

B.B. and S.M. supervised the study, they respectively wrote the non-linear optics and EDX part. C.A. did the experiments and wrote the introduction, experimental and XRD part of the letter. L.C. did STEM imaging and wrote STEM-HAADF and ED parts. B.L. did the integrated non-linear optical simulation and wrote the simulation part.

All authors have given approval to the final version of the manuscript.

**Notes**

The authors declare no competing financial interest.

ACKNOWLEDGMENT

The PhD of Adrien Clavel is funded by École Doctorale I-MEP²

The TEM facility JEOL NEOARM at CNRS Institut Neel was co-financed by the European Union under the European Regional Development Fund (ERDF, contract no. RA0023813).

The authors acknowledge the work and thanks M.S. CAVALAGLIO Sébastien that prepared the TEM sample shown in this work.

REFERENCES

(1) Bencheikh, K.; Cenni, M. F. B.; Oudot, E.; Boutou, V.; Félix, C.; Prades, J. C.; Vernay, A.; Bertrand, J.; Bassignot, F.; Chauvet, M.; Bussières, F.; Zbinden, H.; Levenson, A.; Boulanger, B. Demonstrating Quantum Properties of Triple Photons Generated by $\chi^3$ Processes. *Eur. Phys. J. D* **2022**, *76* (10), 186. https://doi.org/10.1140/epjd/s10053-022-00514-3.

(2) Bierlein, J. D.; Ferretti, A.; Brixner, L. H.; Hsu, W. Y. Fabrication and Characterization of Optical Waveguides in KTiOPO$_4$. In *Optical Fiber Communication*; Optica Publishing Group, 1987; p PDP5. https://doi.org/10.1364/OFC.1987.PDP5.

(3) Wang, F.-X.; Chen, F.; Wang, X.-L.; Lu, Q.-M.; Wang, K.-M.; Shen, D.-Y.; Ma, H.-J.; Nie, R. Fabrication of Optical Waveguides in KTiOAsO$_4$ by He or Si Ion Implantation. *Nucl. Instrum. Methods Phys. Res. Sect. B Beam Interact. Mater. At.* **2004**, *215* (3), 389–393. https://doi.org/10.1016/j.nimb.2003.09.011.

(4) Vernay, A.; Boutou, V.; Félix, C.; Jegouso, D.; Bassignot, F.; Chauvet, M.; Boulanger, B. Birefringence Phase-Matched Direct Third-Harmonic Generation in a Ridge Optical Waveguide Based on a KTiOPO$_4$ Single Crystal. *Opt. Express* **2021**, *29* (14), 22266–22274. https://doi.org/10.1364/oe.432636.

(5) Christen, H. M.; Eres, G. Recent Advances in Pulsed-Laser Deposition of Complex Oxides. *J. Phys. Condens. Matter* **2008**, *20* (26). https://doi.org/10.1088/0953-8984/20/26/264005.

(6) Shepelin, N. A.; Tehrani, Z. P.; Ohannessian, N.; Schneider, C. W.; Pergolesi, D.; Lippert, T. A Practical Guide to Pulsed Laser Deposition. *Chem Soc Rev* **2023**, *52* (7), 2294–2321. https://doi.org/10.1039/D2CS00938B.

(7) Kilburger, S.; Chety, R.; Millon, E.; Di Bin, Ph.; Di Bin, C.; Boulle, A.; Guinebretière, R. Growth of LiNbO3 Thin Films on Sapphire by Pulsed-Laser Deposition for Electro-Optic Modulators. *Appl. Surf. Sci.* **2007**, *253* (19), 8263–8267. https://doi.org/10.1016/j.apsusc.2007.02.112.

(8) Deng, Y.; Du, Y. L.; Zhang, M. S.; Han, J. H.; Yin, Z. Nonlinear Optical Properties in SrTiO3 Thin Films by Pulsed Laser Deposition. *Solid State Commun.* **2005**, *135* (4), 221–225. https://doi.org/10.1016/j.ssc.2005.04.031.

(9) Vasa, N. J.; Hata, Y.; Yoshitake, T.; Yokoyama, S. Preparation of KTiOPO4 Thin Films on Different Substrates by Pulsed Laser Deposition. *Int. J. Adv. Manuf. Technol.* **2008**, *38* (5), 600–604. https://doi.org/10.1007/s00170-007-1040-x.

(10) Kumar, V.; Singh, S. K.; Sharma, H.; Kumar, S.; Banerjee, M. K.; Vij, A. Investigation of Structural and Optical Properties of ZnO Thin Films of Different Thickness Grown by Pulsed Laser Deposition Method. *Phys. B Condens. Matter* **2019**, *552*, 221–226. https://doi.org/10.1016/j.physb.2018.10.004.

(11) Liu, Z. G.; Liu, J. M.; Ming, N. B.; Wang, J. Y.; Liu, Y. G.; Jiang, M. H. Epitaxial Growth of RbTiOPO4 Films on KTiOPO4 Substrates by Excimer Laser Ablation Technique. *J. Appl. Phys.* **1994**, *76* (12), 8215–8217. https://doi.org/10.1063/1.357884.

(12) Salaün, M.; Thiam, A.; Kodjikian, S.; Boulanger, B. Growth and Characterization of Rubidium Titanyl Phosphate Thin Films by Pulsed Laser Deposition. *Materialia* **2024**, *34*, 102068. https://doi.org/10.1016/j.mtla.2024.102068.

(13) Bierlein, J. D.; Vanherzeele, H.; Ballman, A. A. Linear and Nonlinear Optical Properties of Flux-grown KTiOAsO$_4$. *Appl. Phys. Lett.* **1989**, *54* (9), 783–785. https://doi.org/10.1063/1.101552.

(14) Boulanger, B.; Fève, J. P.; Marnier, G.; Ménaert, B. Methodology for Optical Studies of Nonlinear Crystals: Application to the Isomorph Family KTiOPO$_4$, KTiOAsO$_4$, RbTiOAsO$_4$ and CsTiOAsO$_4$. *Pure Appl. Opt. J. Eur. Opt. Soc. Part A* **1998**, *7* (2), 239–256. https://doi.org/10.1088/0963-9659/7/2/014.

(15) Wagner, H. P.; Wittmann, S.; Schmitzer, H.; Stanzl, H. Phase Matched Second Harmonic Generation Using Thin Film ZnTe Optical Waveguides. *J. Appl. Phys.* **1995**, *77* (8), 3637–3640. https://doi.org/10.1063/1.358600.

(16) Boulanger, B.; Fejer, M. M.; Blachman, R.; Bordui, P. F. Study of KTiOPO$_4$ Gray-tracking at 1064, 532, and 355 Nm. *Appl. Phys. Lett.* **1994**, *65* (19), 2401–2403. https://doi.org/10.1063/1.112688.

(17) Boulanger, B.; Rousseau, I.; Feve, J. P.; Maglione, M.; Menaert, B.; Marnier, G. Optical Studies of Laser-Induced Gray-Tracking in KTP. *IEEE J. Quantum Electron.* **1999**, *35* (3), 281–286. https://doi.org/10.1109/3.748831.